\documentclass[sn-mathphys-ay]{sn-jnl}% Math and Physical Sciences Author Year Reference Style
\usepackage{graphicx}           % 图形包
\usepackage{url}                % URL 支持
\usepackage{multirow}           % 支持多行单元格
\usepackage{amsmath,amssymb,amsfonts} % 数学符号、字体
\usepackage{amsthm}             % 定理相关
\usepackage{mathrsfs}           % 花体字体
\usepackage[title]{appendix}    % 附录
\usepackage{pgfplotstable}      % 表格绘图
\usepackage{natbib}             % 引用
\usepackage{float}              % 浮动体支持
\usepackage{booktabs}           % 精美的表格
\usepackage{algorithm}          % 算法
\usepackage{algorithmicx}       % 算法包
\usepackage{algpseudocode}      % 算法伪代码
\usepackage{listings}           % 程序代码支持
\usepackage{svg}                % SVG 图形支持
\usepackage{csvsimple}          % 读取 CSV 文件
\usepackage{appendix}           % 附录包

\theoremstyle{thmstyleone}%
\theoremstyle{thmstyletwo}%

\theoremstyle{thmstylethree}%

\begin{document}

\title[Article Title]{Dynamic ETF Portfolio Optimization Using enhanced Transformer-Based Models for Covariance and Semi-Covariance Prediction(Working Version)}

\author[1]{\fnm{Jiahao} \sur{Zhu}}\email{viktor95@terpmail.umd.edu}
\author[2]{\fnm{Hengzhi} \sur{Wu}}\email{ stephenbutters@ucla.edu}
\affil[1]{\orgdiv{Smith Bussiness School}, \orgname{University of Maryland}, \orgaddress{\street{4340 Van Munching Hall}, \city{University of Maryland College Park}, \postcode{20783}, \state{MD}, \country{U.S.}}}
\affil[2]{\orgdiv{School of Computer Science}, \orgname{University of California,Los Angeles}, \orgaddress{\street{405 Hilgard Avenue}, \city{Los Angeles}, \postcode{90095}, \state{CA}, \country{USA}}}

\newpage

\abstract{This study explores the use of Transformer-based models to predict both covariance and semi-covariance matrices for ETF portfolio optimization. Traditional portfolio optimization techniques often rely on static covariance estimates or impose strict model assumptions, which may fail to capture the dynamic and non-linear nature of market fluctuations. In contrast, our approach leverages the power of Transformer models—specifically Autoformer, Informer, and Reformer—to generate adaptive, real-time predictions of asset covariances, with a particular focus on the semi-covariance matrix to account for downside risk. The semi-covariance matrix, which emphasizes negative correlations between assets, offers a more nuanced approach to risk management compared to traditional methods that treat all volatility equally.

Through a series of experiments, we demonstrate that Transformer-based predictions of both covariance and semi-covariance significantly enhance portfolio performance. Our results show that portfolios optimized using the semi-covariance matrix outperform those optimized with the standard covariance matrix, particularly in volatile market conditions. Moreover, the use of the Sortino ratio—a risk-adjusted performance metric that focuses on downside risk—further validates the effectiveness of our approach in managing risk while maximizing returns.

These findings have important implications for asset managers and investors, offering a dynamic, data-driven framework for portfolio construction that adapts more effectively to shifting market conditions. By integrating Transformer-based models with the semi-covariance matrix for improved risk management, this research contributes to the growing field of machine learning in finance and provides valuable insights for optimizing ETF portfolios.}




\maketitle
\newpage
\section{Introduction}\label{sec1}

\indent

Extensive research has shown that diversification is a key strategy in portfolio optimization, effectively reducing risk and improving overall performance. However, two fundamental challenges continue to hinder the practical application of these principles, especially when employing the classical Capital Asset Pricing Model (CAPM) framework introduced by Fama and French \citep{FAMA19933}. The first challenge lies in the use of mean-variance analysis, which relies on the covariance matrix for volatility estimation. While various studies suggest that asset correlations tend to be more stable than asset prices, modeling and estimating the covariance matrix using traditional linear methods remains problematic. This is primarily due to the high dimensionality of covariance models, where changes in multidimensional spaces are not necessarily linear.

Additionally, the intuition behind equating variance with risk is debatable. Many investors consider upside fluctuations as potential opportunities, not risks, while using variance as a measure of risk neglects this distinction. Despite this, variance continues to be widely used in portfolio optimization due to its positive semi-definiteness \citep{b76ccb64-7fa5-32f9-a4fe-72298146be7d}, which makes it easier to apply in optimization algorithms.

A more intuitive approach, however, would focus exclusively on downside fluctuations. Recognizing the limitations of the covariance matrix, Estrada \citep{semivariance} proposed using semi-covariance as an alternative that better aligns with risk-averse investors' concerns. Semi-covariance, which captures only the negative deviations between asset returns, provides a more accurate measure of downside risk. However, the semi-covariance matrix faces similar challenges as the covariance matrix in terms of estimation and modeling, particularly because, unlike the covariance matrix, it is not positive semi-definite, complicating its use in optimization.

To address these challenges, this article proposes a new approach for constructing a more intuitive and risk-averse portfolio by focusing on Exchange-Traded Funds (ETFs) for increased diversification. Specifically, we substitute the traditional covariance matrix with the semi-covariance matrix in the portfolio optimization process. This adjustment aims to produce portfolios that are better aligned with investors' preferences for minimizing downside risk. Furthermore, this article leverages the power of advanced Transformer-based models to improve the estimation of both covariance and semi-covariance matrices. These models, including Autoformer, Informer, and Reformer, offer a dynamic, adaptive framework for capturing the non-linear and time-varying relationships between asset returns. We then compare the performance of portfolios optimized with these Transformer-based predictions against traditional optimization approaches to assess their effectiveness in improving risk-adjusted returns.
\indent

\section{Related Work}
\indent
Our work builds upon several research strands in finance and machine learning. Specifically, we draw upon the extant literature on CAPM portfolio optimization framework, transformer-based model and attention mechanism framework, and mean-variance optimization framework to develop our methodology. 

\indent
\subsection{The Mean-Variance Analysis and Portfolio Optimization}

\indent
Modeling portfolio volatility is a crucial component across multiple financial domains and has been the focus of numerous academic studies. Volatility serves as a key indicator of uncertainty and is a decisive variable in many investment decisions and portfolio constructions. As proposed by \citep{3c561789-0e17-3c2d-bd8e-66604ddf4002}, such uncertainty is defined as financial risk and plays a vital role in determining the market return of an asset. Previous work in this domain has predominantly concentrated on the estimation of the covariance matrix. For instance, \citep{Markowitz1952} illustrates the equation for calculating portfolio variance as:

\[ \sigma^2_p = \mathbf{w}^T \Sigma \mathbf{w} \]

where \(\mathbf{w}\) is the weight vector indicating the weight of each asset in the portfolio, and \(\Sigma\) is the covariance matrix.

Proposed by \cite{BlackLitterman1992}, the Mean-Variance Analysis framework has become one of the most dominant methods used to evaluate and optimize portfolios, focusing primarily on two aspects: expected return and risk. As mentioned above, risk is defined as the variance of a portfolio, while expected return is the average return that an investor anticipates from the portfolio. It is calculated as the weighted average of the expected returns of the individual assets, with the weights being the proportion of each asset in the portfolio. The goal of portfolio optimization is to maximize the expected return for a given level of risk or to minimize the risk for a given level of return. This process involves constructing the efficient frontier and finding the optimal portfolio according to the investor's risk preference.

To further evaluate the performance in constructing the forecasting method of the covariance and semi-covariance matrix, and considering the regulation in the Chinese stock market that greatly limits the ability to create short positions \citep{ChineseRegulations}, this article aims to optimize our portfolio toward a Minimum Variance Portfolio, which is the portfolio with the lowest risk (variance) among all possible portfolios. This portfolio is particularly suitable for risk-averse investors \citep{MinimumVariance}. Therefore, utilizing the Lagrange multiplier method, the optimization problem becomes minimizing:

\[ \sigma_p^2 = \mathbf{w}^T \mathbf{\Sigma} \mathbf{w} \]

subject to the constraint:

\[ \sum_{i=1}^{n} w_i = 1 \]

Thus, the Lagrange multiplier equation can be written as:

\[ \mathcal{L}(\mathbf{w}, \lambda) = \mathbf{w}^T \mathbf{\Sigma} \mathbf{w} + \lambda (1 - \mathbf{w}^T \mathbf{1}) \]

The final equation for the weight of each asset is:

\[ \mathbf{w} = \frac{\mathbf{\Sigma}^{-1} \mathbf{1}}{\mathbf{1}^T \mathbf{\Sigma}^{-1} \mathbf{1}} \]

\indent

\subsection{Portfolio Diversification and Exchange Traded Fund}

\indent

Since Sharpe (1964) \citep{Sharpe1964} developed the Capital Asset Pricing Model (CAPM) building on Modern Portfolio Theory (MPT), the model has been instrumental in explaining how diversification impacts expected returns and risk. Numerous empirical studies \citep{DiversificationEmpiricalStudies}\citep{FamaEMH} have demonstrated the superiority of diversification in minimizing portfolio risk. The advent of Exchange-Traded Funds (ETFs) has revolutionized portfolio diversification by providing easy access to a wide range of asset classes, sectors, and geographic regions.

An Exchange-Traded Fund (ETF) is a type of investment fund and exchange-traded product, meaning that they are traded on stock exchanges. ETFs typically hold a diversified portfolio of assets, which can include a wide range of stocks, bonds, commodities, or other assets. This diversification helps reduce risk compared to holding individual securities.

According to Eugene Fama's Efficient Market Hypothesis (EMH) \citep{FamaEMH}, asset prices fully reflect all available information in the market. This theory underpins the idea that passive investment vehicles like ETFs, which aim to replicate market indices, are effective tools for diversification. By using sector-specific ETFs, investors can diversify across different sectors of the economy and adjust their exposure based on economic cycles or sector performance expectations \citep{SectorETFs}.

The Arbitrage Pricing Theory (APT) \citep{APT1976}, developed as an alternative to CAPM, asserts that multiple factors, rather than a single market factor, influence asset returns. This theory supports the use of diversified portfolios, including ETFs, to capture various risk premiums.
\indent

\subsection{Covariance Estimation and Forecasting}

\indent
Covariance estimation and forecasting are crucial components in financial modeling, particularly in the context of portfolio management, risk assessment, and asset pricing. Since Markowitz \citep{Markowitz1952} laid the foundation for Modern Portfolio Theory (MPT), emphasizing the importance of covariance in portfolio optimization, a plethora of research has been dedicated to the estimation and modeling of the covariance matrix. The sample covariance matrix is the most straightforward method, calculated directly from historical return data. Given $n$ observations of $p$ assets, let $X$ be an $n \times p$ matrix of asset returns. The sample covariance matrix $S$ is given by:

$S = \frac{1}{n-1} (X - \bar{X})^T (X - \bar{X})$

where $\bar{X}$
  is the matrix of mean returns. However, this method makes no assumptions about data distribution beyond having sufficient historical data, making it noisy and unstable, especially with a small sample size relative to the number of assets. Consequently, it may not be well-conditioned in high-dimensional settings.

Given the limitations of the sample covariance matrix, some efforts have been dedicated to improving it by combining it with a structured estimator (like the identity matrix) to reduce estimation error. The famous Ledoit-Wolf Shrinkage \citep{LedoitWolf2004} is one of the most popular shrinkage methods, which shrinks the sample covariance matrix towards a scaled identity matrix:

$\Sigma_{\text{shrinkage}} = \lambda T + (1 - \lambda) S$

where $T$ is the target matrix (often the identity matrix), $S$ is the sample covariance matrix, and $\lambda$ is the shrinkage intensity. This shrinkage method reduces estimation error, but the choice of shrinkage target and intensity parameter can be somewhat arbitrary and require tuning \citep{LedoitWolf2004}.

There are also various methods for modeling the covariance matrix, such as the Multivariate GARCH models including the Diagonal VECH Model \citep{Bollerslev1988}, BEKK Model \citep{Engle1995}, and Dynamic Conditional Correlation (DCC) \citep{Engle2002}. High-dimensional and regularization techniques such as Graphical Lasso \citep{Friedman2008}, and Thresholding Methods \citep{Bickel2008} have also been explored. However, they all have their own limitations. The Multivariate GARCH models require strong statistical assumptions and can be sensitive to model specification \citep{Bollerslev1988}. High-dimensional and regularization techniques may introduce bias if the underlying true covariance is not sparse \citep{Bickel2008}.

In view of the limitations posed by different sets of methods, some researchers have turned to the field of trending machine learning techniques. Bollerslev \citep{Bollerslev2018} explores methods for improving volatility forecasting by using machine learning techniques to model dynamic covariances between financial assets. Heaton \citep{Heaton2016} discusses the use of deep learning models, including auto-encoders and recurrent neural networks, for modeling the covariance structure of financial time series and their application in portfolio management. Others have developed methods utilizing LSTM networks, random forests to model consumer credit risk. However, the actual use of deep learning networks in covariance matrix and portfolio optimization remains under-explored, and none have yet utilized the prestigious transformer-based models in this domain. This is partially due to the limitation of large models in explaining the intuition of the result (black boxes) \citep{Rudin2019}, and the complexity of balancing between imposing various constraints and achieving prediction accuracy in the task of modeling covariance by deep-learning methods.

\indent
\subsection{Semicovariance and Mean-Semivariance Optimization}

Although Markowitz \citep{Markowitz1952}, Sharpe \citep{Sharpe1964}, and Fama-French \citep{Fama1992} have laid the foundation of the Capital Asset Pricing Model (CAPM) framework, which utilizes the covariance matrix as a key element in volatility estimation and treats the variance of return as a measure of risk, this traditional approach has its limitations. For many investors, upside fluctuations are not perceived as risks, while using variance as a measure neglects this fact. Despite this limitation, variance is widely employed in portfolio optimization due to its positive semi-definiteness characteristic \citep{PSD_in_covariance}, which facilitates the optimization process.

A more intuitive approach would be to focus exclusively on downside fluctuations. Recognizing the limitations of relying solely on the covariance matrix in portfolio optimization, Estrada \citep{Estrada2002} proposed the use of semi-covariance as a more intuitive alternative. However, the semi-covariance matrix encounters challenges similar to those of the covariance matrix concerning estimation and modeling. Moreover, unlike the covariance matrix, the semi-covariance matrix, as defined by Estrada, is not positive semi-definite. If one substitutes the covariance matrix with the semi-covariance matrix under the framework of CAPM, it would inevitably complicate the optimization process further compared to that which utilizes the covariance matrix.

Semi-covariance measures the covariance of asset returns that fall below a certain threshold, typically the mean or a target return, and is defined as:

$\text{SemiCov}(X, Y) = \mathbb{E}[\min(X - \mu_X, 0) \min(Y - \mu_Y, 0)]$

where $X$ and $Y$ are asset returns and $\mu_X$
  and $\mu_Y$
  are their respective means. This focus on downside risk aligns more closely with the risk aversion of investors who are primarily concerned with losses rather than gains.

Despite the theoretical appeal of semi-covariance, practical challenges arise. The estimation of the semi-covariance matrix can be as noisy and unstable as the sample covariance matrix, especially with limited historical data. Additionally, the lack of positive semi-definiteness complicates optimization algorithms, which rely on this property to ensure that the optimization problem is convex \citep{Boyd2004} and thus solvable using efficient numerical methods \citep{Nesterov1994}.

In portfolio optimization, convexity is crucial because it guarantees that the optimization problem has a unique global minimum and that numerical techniques can efficiently find this minimum. Convex optimization problems are well-understood and have robust solution methods, such as quadratic programming, which benefit from the positive semi-definite nature of the covariance matrix.

However, when using the semi-covariance matrix, the absence of positive semi-definiteness introduces non-convexity into the optimization problem. This makes the optimization landscape much more complex, with the potential for multiple local minima, and requires more sophisticated and computationally intensive algorithms to solve.

In conclusion, while the semi-covariance matrix provides a potentially more intuitive measure of risk by focusing on downside fluctuations, its practical application in portfolio optimization is hampered by estimation challenges and mathematical complexities. These issues highlight the ongoing need for robust methods that can accurately capture risk while being computationally feasible for practical portfolio management.

\indent
\subsection{Transformer-based Model and Numerical Forecasting}
\indent
Extensive research has been conducted on leveraging Transformer-based models for predicting numerical time-series data. Studies consistently underscore the advantages of these models over traditional time-series approaches \citep{chu2024unettsf}, highlighting the enhanced ability of deep learning networks like Transformers to adeptly capture the non-linear dynamics inherent in complex datasets. Specifically, researchers have explored the application of Transformers in forecasting financial data, including stock prices, market trends, and volatility \citep{DBLP:journals/ngc/HiranoIS22}\citep{DBLP:journals/eswa/WangCZZ22}. Financial datasets are often characterized by high volatility, noise, and non-stationarity—traits that present substantial challenges for linear modeling. Transformers, with their advanced architecture, offer a compelling alternative, effectively handling these intricate data attributes \citep{DBLP:conf/aiccsa/AlissaA22}.

Research employing Transformer-based models for financial analysis can be categorized into two main areas. The first focuses on innovations within the attention mechanisms of the Transformer’s encoder and decoder modules. For instance, the Reformer model \citep{DBLP:conf/iclr/KitaevKL20} introduces an enhancement by replacing the traditional dot-product attention mechanism with locality-sensitive hashing, while studies such as \citep{DBLP:journals/corr/abs-2012-07436} propose the Informer model, designed to efficiently process the lengthy sequences common in financial data. Building on the Reformer’s principles, the Informer employs a Prob-Sparse self-attention mechanism, offering a sophisticated adaptation to the unique demands of financial time-series analysis. Additionally, \citep{DBLP:journals/apin/MaZZKB23} combine Transformers with convolutional neural networks (CNNs) and long short-term memory (LSTM) models to address limitations in traditional models when capturing spatial and spatial-temporal features between variables.

The second research area centers on the pre-processing or decomposition of time-series data before their integration into the Transformer’s attention mechanism. Notable examples include the Autoformer \citep{DBLP:journals/corr/abs-2106-13008}, which employs a decomposition strategy to separate time-series data into trend, seasonal, and cyclic components. Similarly, the FEDformer \citep{DBLP:journals/corr/abs-2201-12740} leverages the observation that many time series exhibit sparse representations in bases such as the Fourier transform, using this insight to develop a frequency-enhanced Transformer model.

Despite the extensive exploration of Transformer-based models and their proven superiority over traditional linear approaches in financial analysis, none have yet applied these models to predict the covariance and semi-covariance matrices of financial data. This gap may stem from the complexity involved in ensuring the deep-learning network’s output remains a positive semi-definite (PSD) matrix while maintaining prediction accuracy \citep{DBLP:journals/corr/abs-1902-11189}. Addressing this challenge is a key aim of the present study.
\indent

\section{Methodology}

\begin{figure*}[ht!]
\centerline{\includegraphics[width=1\textwidth, keepaspectratio]{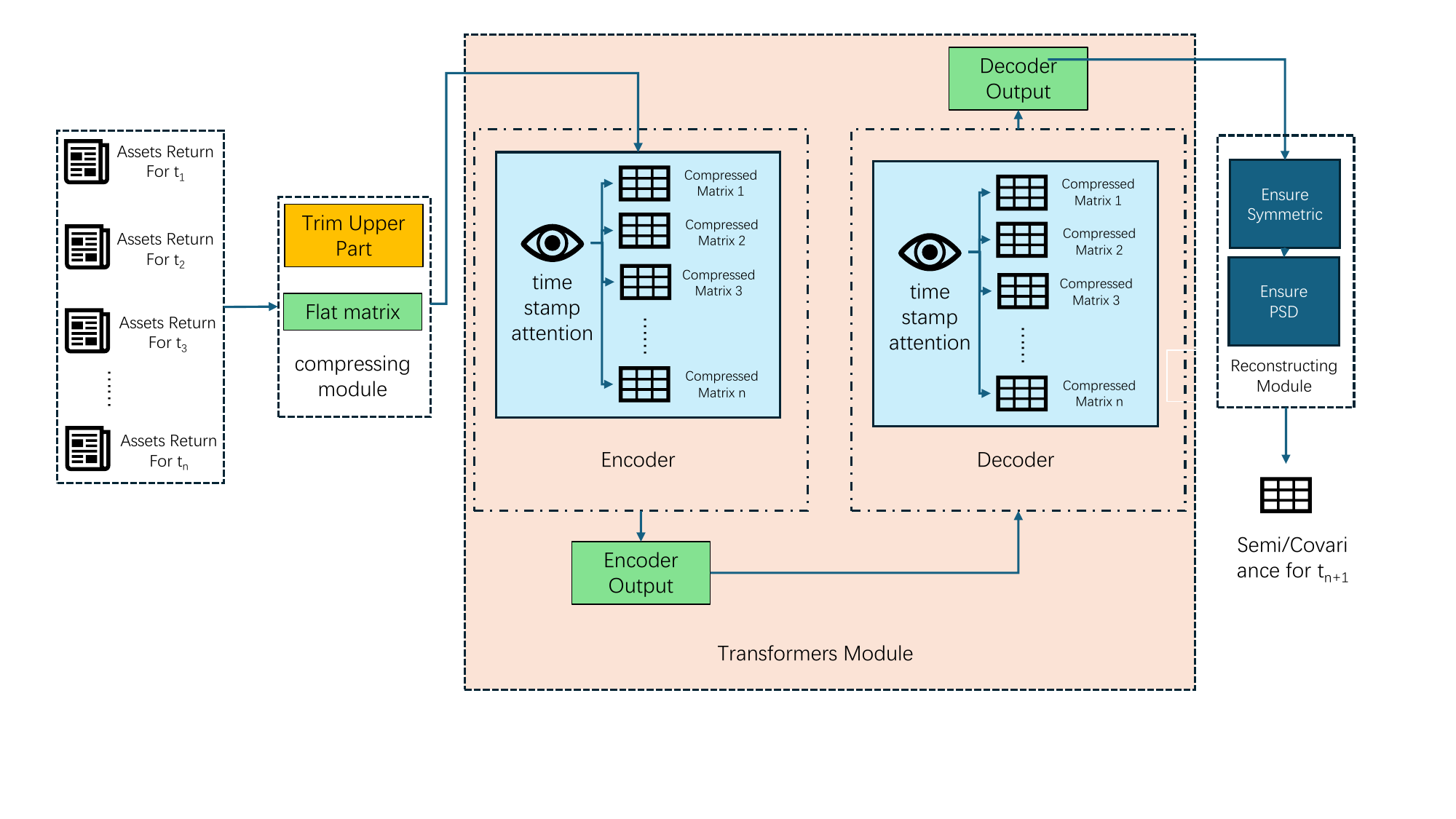}}
\caption{ Architecture Overview.}
\label{fig1}
\end{figure*}
To investigate the prediction accuracy of the Transformer-based methodology for modeling covariance and semi-covariance, as well as its performance in portfolio optimization, this article proposes the methodological framework outlined below. The proposed method will be evaluated from two perspectives: first, by comparing it to the baseline using sample covariance estimation, and second, by comparing it to the baseline using the semi-covariance matrix. For an overview of the overall model structure, please refer to Figure \ref{fig1}.

\subsection{Compressing Module}
In this study, we propose a novel approach for predicting the covariance matrix of financial data using Transformer-based models. A key challenge in financial forecasting, particularly when dealing with covariance matrices, is the high dimensionality and complex dependencies between assets. To address this, we introduce a dimensionality reduction technique that compresses the last two dimensions of the covariance matrix into a single dimension, effectively reducing the complexity of the input data while preserving the essential relationships between asset pairs. Specifically, the covariance matrix $\Sigma \in \mathbb{R}^{n \times n}$
 , which represents the pairwise covariances between $n$ assets, is symmetric, meaning $\Sigma = \Sigma^T$
 . The matrix has 
 $\frac{n(n+1)}{2}$
  unique elements, as the off-diagonal elements (i.e., the covariances between distinct assets) mirror each other.

To reduce the dimensionality of the input, we flatten the lower triangular part (or equivalently, the upper triangular part) of the covariance matrix into a vector. Mathematically, this operation can be expressed as:

$$
\mathbf{v} = \text{vec}(\text{Tri}(\Sigma)) \in \mathbb{R}^{\frac{n(n+1)}{2}}
$$

where $\text{Tri}(\Sigma)$ denotes the lower triangular part of $\Sigma$, and $\text{vec}(\cdot)$ is the vectorization operator that stacks the elements of the matrix into a single column vector. The resulting vector $\mathbf{v}$ has a size of $\frac{n(n+1)}{2}$
 , which represents the compressed form of the covariance matrix.

Once the dimensionality is reduced, this vector $\mathbf{v}$ is fed into the Transformer architecture. 

\subsection{Time steps Attention Mechanism}

The Transformer model, originally designed for natural language processing tasks, has emerged as a powerful tool for time-series prediction, thanks to its unique attention mechanism. Traditional time-series models, such as ARIMA or Exponential Smoothing, typically rely on fixed dependencies or assumptions about temporal structures. In contrast, Transformer models, by employing self-attention mechanisms, can learn dynamic dependencies across time steps, regardless of their position in the sequence. This capability makes them particularly effective for modeling complex, long-range dependencies that are common in real-world time-series data.

The self-attention mechanism allows the model to weigh the importance of different time steps when making predictions, enabling it to focus on the most relevant historical data points at any given time. In time-series forecasting, this means that the Transformer model can capture both local patterns (e.g., daily fluctuations in ) and global trends (e.g., long-term economic shifts), irrespective of their temporal distance from the current time point. This flexibility is a key advantage over traditional methods that may struggle to account for non-linear relationships and long-range dependencies.

convinced by the work of \citep{Bollerslev1986, Engle2002} that the correlation between financial assets has a relatively stable structure through time. Our model uses a time stamp attention mechanism to predict the future covariance matrix. The self-attention mechanism in the Transformer model \citep{Vaswani2017} computes attention scores for each asset's correlation relative to all other assets, enabling the model to effectively capture how the returns of one asset affect another over time.

The self-attention operation can be formulated as:

$$
\text{Attention}(Q, K, V) = \text{softmax}\left(\frac{QK^T}{\sqrt{d_k}}\right) V
$$

where:

$Q$, $K$, and $V$ are the query, key, and value matrices, derived from the input covariance time-series. These matrices represent the covariance data after the transformation from the Compressing module.
$d_k$
  is the dimensionality of the key vectors.
The softmax function ensures that the attention weights sum to 1, making them interpretable as relative importance scores.
By applying self-attention, the Transformer model can dynamically learn which time steps or asset pairs are most important for predicting the covariance matrix at each point in time, allowing for more accurate and adaptive predictions.

While self-attention captures dependencies between time steps or assets, multi-head attention further enhances the model's ability to capture diverse relationships by learning multiple attention patterns simultaneously. In multi-head attention, multiple sets of query, key, and value weight matrices are learned, each focusing on different aspects of the financial data. The outputs from these attention heads are concatenated and linearly transformed to form the final prediction. This mechanism allows the model to capture different types of temporal or asset-related dependencies, such as short-term fluctuations, long-term trends, or sector-specific correlations.

The multi-head attention mechanism can be expressed as:

$$
\text{MultiHead}(Q, K, V) = \text{Concat}(\text{head}_1, \text{head}_2, \dots, \text{head}_h) W^O
$$

where each $\text{head}_i$
  is the attention output from the $i$-th head, and $W^O$
  is the output transformation matrix. The ability of multi-head attention to capture various patterns of relationships across assets over time makes it particularly well-suited for the task of predicting complex covariance matrices, which involve multi-dimensional interactions and non-linear correlations.

When applied to covariance matrix prediction, the Transformer model operates on a transformed representation of the covariance matrix. First, the lower triangular part of the covariance matrix is extracted, capturing the unique pairwise relationships between assets. This lower triangular matrix is then processed through a compression module, which reduces its dimensionality before feeding it into the Transformer. By doing so, the model is able to focus on essential inter-asset relationships without being overwhelmed by unnecessary complexity. The Transformer is then trained to predict the future covariance structure, learning from past relationships and adapting its attention mechanism to capture both the temporal dependencies and inter-asset correlations inherent in the financial data. This predictive capability is vital for portfolio optimization, where accurate estimates of asset covariances are essential for constructing portfolios that maximize risk-adjusted returns.

\subsection{Covariance Reconstructing Module}

The Transformer model, with its self-attention mechanism, is well-suited for capturing the long-range dependencies and non-linear relationships between the elements of the vector, which correspond to the pairwise covariances of the assets. After training, the Transformer model predicts the future covariance vector, which is then reshaped back into the original covariance matrix format by reversing the vectorization process:

$$
\hat{\Sigma} = \text{Unvec}(\hat{\mathbf{v}}) \in \mathbb{R}^{n \times n}
$$
 
where $\hat{\mathbf{v}}$
  is the predicted covariance vector, and $\text{Unvec}(\cdot)$ reconstructs the full matrix from the vector. This reconstructed covariance matrix $\hat{\Sigma}$
  can then be used for downstream applications, such as portfolio optimization, to make informed investment decisions based on the predicted asset correlations.

\subsection{Regulated Network Optimization }
\indent
Following the extraction of the lower triangular covariance matrix, application of a compression module, and processing through the Transformer model, a critical step in our methodology is to ensure that the output matrix approximates a positive semi-definite (PSD) matrix. A PSD covariance matrix is necessary for applications like portfolio optimization, where negative eigenvalues can lead to invalid or unstable results \citep{LedoitWolf2004, michaud1989expected}. While a regularized loss function can encourage the predicted covariance matrix to approximate PSD, it does not necessarily guarantee strict positive semi-definiteness due to the inherent complexity of financial data and the optimization process \citep{balakrishnan2020estimating}.

Our loss function includes several components. The primary loss, $\mathcal{L}_{\text{reg}}$
 , evaluates the difference between the predicted and true covariance matrices using a mean squared error criterion \citep{bollerslev1990modelling}. For a given batch, the model processes past input data $X_{\text{past}}$ and returns predictions for the covariance matrix, denoted as $\hat{\Sigma}_{\text{pred}} $. The loss is then calculated based on the mode of operation:

$$
\mathcal{L}_{\text{MSE}} = \| \hat{\Sigma}_{\text{pred}} - \Sigma_{\text{true}} \|^2
$$
 
where $\Sigma_{\text{true}}$
  is the true covariance matrix.

To further ensure that the predicted covariance matrix approximates the PSD property \citep{goodfellow2016deep}., we add two regularization terms. The first term, $\mathcal{L}_{\text{sym}}$
 , penalizes asymmetry by calculating the mean absolute difference between the predicted matrix and its transpose:

$$
\mathcal{L}_{\text{sym}} = \frac{1}{n^2} \sum_{i,j} \left| \hat{\Sigma}_{\text{pred}, ij} - \hat{\Sigma}_{\text{pred}, ji} \right|
$$
 
where $n$ is the dimensionality of the matrix \citep{higham1988computing}. This loss encourages the matrix to remain symmetric.

The second regularization term, $\mathcal{L}_{\text{PSD}}$
 , directly promotes positive definiteness by penalizing any negative real components in the matrix's eigenvalues. We compute the eigenvalues $\lambda_i$
  of the predicted average covariance matrix $\hat{\Sigma}_{\text{pred}}$, and apply a clamp function to shift any negative real components to zero:
$$
\mathcal{L}_{\text{PSD}} = \frac{1}{n} \sum_{i} \max(0, -\operatorname{Re}(\lambda_i))
$$
where $
\operatorname{Re}(\lambda_i)
$denotes the real part of the eigenvalue \citep{bai2010spectral}. This term encourages the matrix to be closer to PSD by penalizing the presence of negative eigenvalues.

The final training loss, $\mathcal{L}_{\text{total}}$
 , combines these terms, weighted by a penalty factor $\alpha$ that controls the strength of the regularization:

$$
\mathcal{L}_{\text{total}} = \mathcal{L}_{\text{single}} + \alpha (\mathcal{L}_{\text{sym}} + \mathcal{L}_{\text{PSD}})
$$

This aggregated loss is back-propagated through the model, allowing it to iteratively refine its predictions to produce covariance matrices that approximate the PSD property, improving their stability and suitability for downstream financial applications such as portfolio optimization.
\indent
\subsection{Nearest PSD Approximation}

To guarantee that the predicted covariance matrix is positive semi-definite (PSD), we implement a post-processing step on the output of the Autoformer model. This step ensures symmetry, maintains diagonal positivity, and approximates the PSD property of the covariance matrix.

First, we symmetrize the output matrix to ensure it is symmetric, which is a necessary property for covariance matrices. Given the output matrix \( \hat{\Sigma}_{\text{pred}} \) from the Autoformer model, we compute its transpose \( \hat{\Sigma}_{\text{pred}}^T \) and symmetrize the matrix as:
\[
\hat{\Sigma}_{\text{pred}}^{\text{sym}} = \frac{1}{2} \left( \hat{\Sigma}_{\text{pred}} + \hat{\Sigma}_{\text{pred}}^T \right).
\]
This guarantees that \( \hat{\Sigma}_{\text{pred}}^{\text{sym}} \) is symmetric, which is essential for any valid covariance matrix \citep{LedoitWolf2004}.

Next, we adjust the diagonal elements to ensure they are non-negative. Let the diagonal elements of \( \hat{\Sigma}_{\text{pred}}^{\text{sym}} \) be denoted as \( d_i \), where \( d_i \) represents the \( i \)-th diagonal entry. We replace each diagonal entry with its absolute value to ensure positivity, i.e.,
\[
\hat{\Sigma}_{\text{pred}}^{\text{sym}}[i, i] = |d_i|.
\]
This operation ensures that the variances on the diagonal are non-negative, which is a required property for valid covariance matrices \citep{michaud1989expected}.

To further approximate the PSD property, we perform an eigen-decomposition of the symmetrized matrix. The matrix \( \hat{\Sigma}_{\text{pred}}^{\text{sym}} \) is decomposed as:
\[
\hat{\Sigma}_{\text{pred}}^{\text{sym}} = V \Lambda V^T,
\]
where \( V \) is the matrix of eigenvectors and \( \Lambda \) is the diagonal matrix of eigenvalues \( \lambda_i \) of \( \hat{\Sigma}_{\text{pred}}^{\text{sym}} \). To ensure the matrix is positive semi-definite, we clip any negative eigenvalues to zero. In our approach, rather than using absolute values, we set any negative eigenvalue \( \lambda_i \) to zero, i.e.,
\[
\hat{\lambda}_i = \max(\lambda_i, 0).
\]
This procedure avoids negative eigenvalues, which would otherwise make the matrix non-PSD, and ensures that all eigenvalues are non-negative \citep{bai2010spectral}.

Finally, we reconstruct the matrix using the clipped eigenvalues \( \hat{\lambda}_i \). The resulting matrix is:
\[
\hat{\Sigma}_{\text{pred}}^{\text{PSD}} = V \hat{\Lambda} V^T,
\]
where \( \hat{\Lambda} \) is the diagonal matrix of clipped eigenvalues \( \hat{\lambda}_i \). This reconstruction ensures that the matrix is both symmetric and positive semi-definite, making it suitable for downstream financial applications such as portfolio optimization. Through this post-processing step, our model approximates a PSD covariance matrix, improving its stability and utility in financial contexts\citep{higham1988computing}.

\section{Empirical Validation}

\subsection{Data Description and Experimental Setup}

This study begins by constructing an \textit{ETF selection pool} from which the portfolio is formed. The equity portion of the portfolio is based on the \textit{Shenwan Primary Industry Classification}, which categorizes ETFs into industry-specific groups. We selected \textit{industry-specific ETFs} from this classification to ensure sectoral diversification within the portfolio. Additionally, to further broaden the portfolio's exposure and enhance its global diversification, we included major \textit{international stock market index funds}, including the \textit{NASDAQ}, \textit{S\&P 500}, \textit{FTSE 100}, and \textit{MSCI World} indices. To reduce overall portfolio volatility, we incorporated several \textit{money market funds}—assets chosen for their low risk, stable returns, and high liquidity.

Within each asset class (i.e., industry ETFs, international indices, and money market funds), we evaluated the assets based on the following selection criteria:

\begin{enumerate}
    \item \textbf{Longevity}: First, we ranked the ETFs by the number of years they have been operational within each asset class, selecting the top-ranked funds to ensure stability.
    \item \textbf{Expected Returns}: Among the top-ranked ETFs, we selected those with the highest expected returns over the past three years, based on historical performance.
\end{enumerate}

For example, within the \textit{transportation} industry, we first identified the five longest-operating ETFs. From this group, we selected the ETF with the highest expected three-year return as the representative for that sector.

The final list of assets in the ETF portfolio pool, after applying these criteria, includes a diverse set of industry-specific ETFs, international stock indices, and money market funds. A detailed list of these assets, along with their corresponding expected returns and risk profiles, is presented in the appendix.

We used \textit{daily trading data} for each of the selected assets spanning from \textit{March 21, 2022} to \textit{March 21, 2024}, ensuring a sufficiently long sample to capture various market conditions. The data from \textit{March 12, 2022} to \textit{February 12, 2024} was used as the \textit{training set}, while data from \textit{February 12, 2024} to \textit{March 12, 2024} served as the \textit{test set}.

To account for the dynamic nature of financial markets, we employed the \textit{rolling window method} for both the training and testing periods. This method allows the model to be trained on the most recent data and tested on the subsequent period, mimicking real-world portfolio rebalancing. Specifically, for each rebalancing window, we computed the covariance matrix and the semi-covariance matrix to capture both the overall risk (variance) and downside risk (semi-variance) of the ETF Pool for both the training and testing dataset.

The precise parameters used in this process, including the window size and frequency of rebalancing, are not disclosed here due to confidentiality restrictions. The rolling window method ensures that the model reflects the time-varying nature of market conditions, enhancing its practical relevance and robustness.

% Include the table here later
% \begin{table}[htbp]
% \centering
% \begin{tabular}{|c|c|c|}
% \hline
% ETF Name & Expected Return (\%) & Risk Profile (Volatility) \\
% \hline
% ETF 1 & X\% & Y\% \\
% ETF 2 & X\% & Y\% \\
% \vdots & \vdots & \vdots \\
% \hline
% \end{tabular}
% \caption{Detailed List of Assets in the ETF Portfolio Pool.}
% \label{tab:asset_pool}
% \end{table}

% You can include the table here later for asset pool
% \begin{table}[htbp]
% \centering
% \begin{tabular}{|c|c|c|}
% \hline
% ETF Name & Expected Return (\%) & Risk Profile (Volatility) \\
% \hline
% ETF 1 & X\% & Y\% \\
% ETF 2 & X\% & Y\% \\
% \vdots & \vdots & \vdots \\
% \hline
% \end{tabular}
% \caption{Detailed List of Assets in the ETF Portfolio Pool.}
% \label{tab:asset_pool}
% \end{table}

\indent

\subsection{Baseline}

We evaluate the performance of our model from two perspectives.

First, we assess the model's predictive accuracy in estimating both the semi-covariance and covariance matrices. For this, we compare the model's predictions against the sample covariance model, which assumes that the covariance matrix for the upcoming week is the same as the historical covariance matrix from the previous month. This approach is commonly used in traditional portfolio management, where past data is assumed to be a good representation of future market conditions \citep{Markowitz1952}. We measure the accuracy of our model’s estimates by calculating the differences between the predicted and actual covariance matrices over multiple rebalancing windows.

Second, we compare the model’s monthly trading performance with that of models using historical covariance and semi-covariance matrices. The sample covariance models rely on fixed, past covariance and semi-covariance matrices, and their trading strategies are based solely on historical data, without incorporating dynamic, real-time adjustments. In contrast, our model adjusts its portfolio dynamically by incorporating the predicted matrices, which better account for market volatility and downside risk.

As mentioned earlier, this article focuses on minimizing downside fluctuations. To evaluate portfolio performance, we use monthly returns and the Sortino ratio as key performance metrics. The Sortino ratio is particularly suited for this analysis as it focuses on downside risk, which aligns with our goal of minimizing adverse fluctuations. We compare the model’s returns against those of the baseline models over a fixed test period.

The results of these comparisons provide insight into how well our model performs in terms of both predictive accuracy and financial performance relative to traditional approaches based on sampled covariance matrices.

We employed the following transformer-based models for our transformer module:

\begin{itemize}
    \item \textbf{Actual Result}: This compares each result with the one using historical covariance and semi-covariance matrices.
    \item \textbf{Transformer}: This article compares the results of the Transformer model \citep{DBLP:journals/corr/VaswaniSPUJGKP17}.
    \item \textbf{Autoformer}: Based on the Transformer model, \citep{DBLP:journals/corr/abs-2106-13008} add the \textit{Autoformer} module to extract trend, seasonality, and cyclic features as sequences before inputting time-series data into the attention module of the Transformer model.
    \item \textbf{Reformer}: \citep{DBLP:conf/iclr/KitaevKL20} replace dot-product attention in the vanilla Transformer model with one that uses locality-sensitive hashing, changing its complexity from $O(L^2)$ to $O(L\ln{L})$, where $L$ is the length of the input sequences. Furthermore, they use reversible residual layers instead of standard residuals to allow for more sensitive activation in the training process.
    \item \textbf{Informer}: Based on \textit{Reformer}, \citep{DBLP:journals/corr/abs-2012-07436} replace the attention mechanism in Reformer with a Prob-Sparse self-attention mechanism, which achieves $O(L\ln{L})$ in time complexity and memory usage. They also designed a generative-style decoder that predicts long time-series sequences in one forward operation rather than a step-by-step one.
\end{itemize}

\subsection{Training Setup}

This article tunes the hyperparameters based on the validation mean squared error (MSE) to obtain the following values: dropout rate $d = 0.05$, learning rate $\alpha = 0.0001$, and number of epochs = 100. Additionally, the dimension of the node, $d_{\text{model}} = 256$, is used as the number of hidden neurons in our covariance reconstruction module.

\section{Result and Analysis}
\subsection{Portfolio Performance Comparative Analysis}

\begin{figure}[h!]  % 'h!' means to place the figure here, roughly at the location where the code is written
    \centering
    \includegraphics[width=1.2\textwidth]{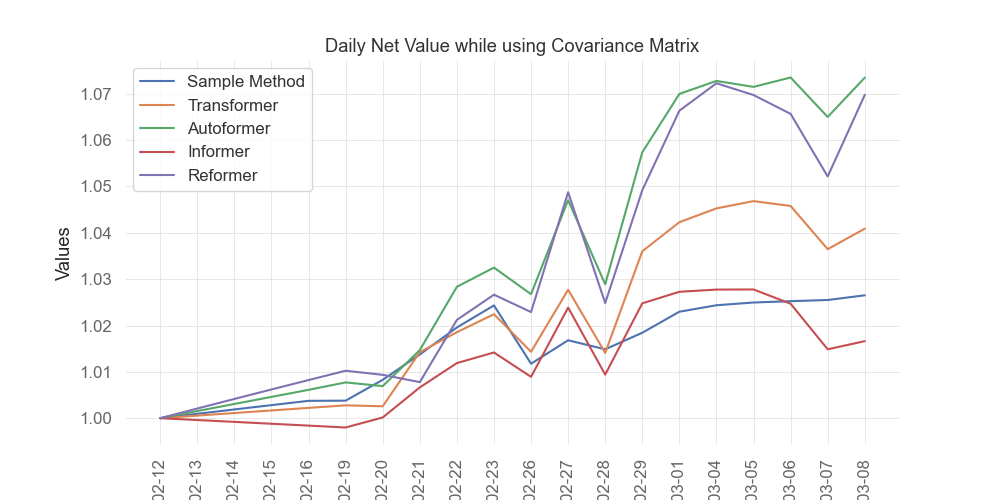}  % Replace 'example-image' with the name of your image file
    \caption{Daily Net value for using covariance modeling.}  % Add a caption
    \label{fig2}  % Add a label for referencing
\end{figure}

First, we evaluate the performance of our portfolio relative to the baseline within our prediction period. The results indicate a consistent improvement across all Transformer-based models in terms of return performance. These results are summarized in Table \ref{table1} and the daily net value using covariance as an optimization matrix is visually represented in Figure \ref{fig2}. 

As shown in Table \ref{table2}, Most Transformer-based models demonstrate steady improvements in portfolio performance compared to the baseline. Figure \ref{fig2} further highlights that all Transformer-based portfolios 
 except Informer outperforms the historical baseline in terms of returns. 
 
 \begin{table}[htbp]
\centering
\begin{tabular}{|l|c|c|c|}
\hline
\textbf{Model} & \textbf{Covariance Matrix MSE} & \textbf{Semi-Covariance Matrix MSE} \\
\hline
\textbf{Sample Method} & 0.00071 & 0.00019\\
\hline
\textbf{Transformer} & 0.00067 & 0.00016\\
\hline
\textbf{Autoformer} & 0.00064 & 0.00015\\
\hline
\textbf{Informer} & 0.00066 & 0.00016\\
\hline
\textbf{Reformer} & 0.00066 & 0.00016\\
\hline
\end{tabular}
\caption{Comparison of Prediction Accuracy for Covariance and Semi-Covariance Matrices using MSE.}
\label{table1}
\end{table}

Upon closer examination of both the actual data and graphical representation, it becomes apparent that the superior performance of our model can be attributed to its more effective handling of market fluctuations. In particular, during major periods of market turbulence, our model consistently increases its premium over the baseline, adapting more effectively to changes in volatility.

\subsection{Covariance and Semi-Covariance Prediction Performance}

While this article observed performance improvements across all Transformer-based models in terms of portfolio returns, these gains may be partly attributed to the greater concentration of the portfolio, which typically results in higher expected returns, or the selected period, which tends to magnify the returns of a more concentrated portfolio in the short run. To explore the relationship between our method and the observed improvement in return performance, we also evaluate our model’s performance in predicting the covariance and semi-covariance matrices.

The final covariance and semi-covariance results for the Transformer-based models and the baseline are summarized in Table \ref{table1}, with Mean Squared Error (MSE) values reported as the average across five runs. To further investigate the model’s behavior.

The results in Table \ref{table1} demonstrate consistent improvements in covariance and semi-covariance prediction across all five Transformer-based models. This performance gain is likely due to the attention mechanism in Transformer models, which allows the model to capture both linear and non-linear transformations within the covariance and semi-covariance matrices over various time windows.

Among the Transformer-based models, Autoformer shows the most significant improvement. This is likely due to Autoformer’s ability to decompose time-series data into trend, seasonal, and cyclical components, which enhances the attention mechanism’s ability to capture and forecast asset correlations over different time horizons. In contrast, Reformer shows more modest gains, which may be attributed to its design trade-offs in terms of model complexity and efficiency. Reformer’s architecture, while more efficient in terms of memory usage, might not capture time-series dependencies as effectively as Autoformer, especially in cases with complex patterns such as seasonal trends.

\subsection{The Covariance and the Semi-Covariance Matrix}

In this section, we extend our previous analysis of predicting covariance matrices using Transformer-based models (Transformer, Autoformer, Informer, and Reformer) and apply the same approach to the semi-covariance matrix. Our core contribution is not only the improved prediction of the semi-covariance matrix but also the novel use of Transformer models to predict both covariance and semi-covariance matrices, which provides a more dynamic, context-aware approach to portfolio optimization.

To assess the effectiveness of both covariance and semi-covariance matrices, we use the Sortino ratio in addition to the final return performance. The Sortino ratio, unlike the Sharpe ratio, focuses on penalizing downside risk rather than overall volatility, making it a more appropriate metric for financial applications where minimizing losses is critical.

\subsubsection{Performance Comparison and Model Implementation}
We applied the same Transformer-based models to predict both covariance and semi-covariance matrices and used these predictions in portfolio optimization. We then compared the performance of portfolios optimized using the covariance and semi-covariance matrices, and evaluated both in terms of final returns and the Sortino ratio, which highlights downside risk-adjusted performance.

\begin{figure}[h!]  % 'h!' means to place the figure here, roughly at the location where the code is written
    \centering
    \includegraphics[width=1.2\textwidth]{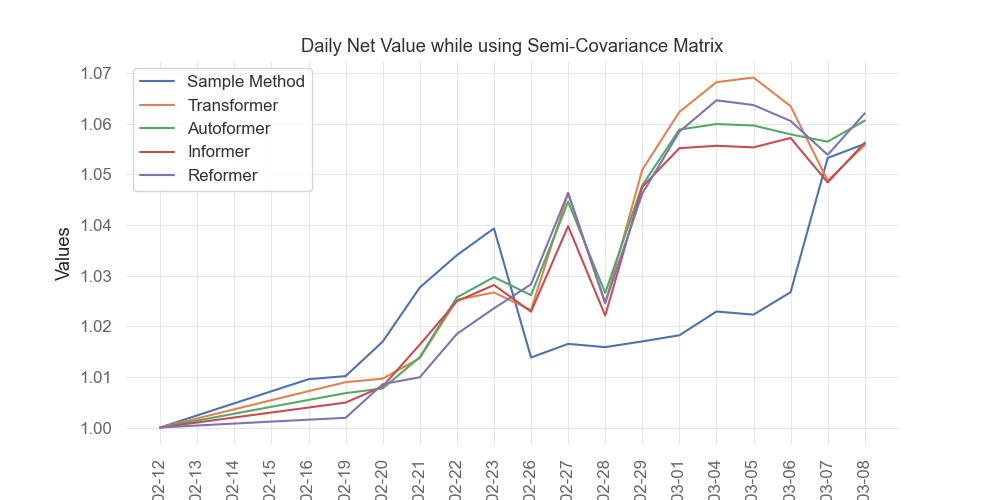}  % Replace 'example-image' with the name of your image file
    \caption{Daily Net value for using semi-covariance modeling.}  % Add a caption
    \label{fig3}  % Add a label for referencing
\end{figure}

Table \ref{table1} provides a summary of the prediction accuracy of the Transformer-Based model base on and Table \ref{table2} shows the final return performance and Sortino ratios for each model using both the covariance and semi-covariance matrices. The results show a clear improvement in performance across all models when the semi-covariance matrix is used. Figure \ref{fig3} illustrates the superior returns and better risk-adjusted performance (via Sortino ratio) achieved by incorporating the semi-covariance matrix.

\begin{table}[ht]
\centering
\resizebox{\textwidth}{!}{ % Resize the table to the width of the page
\begin{tabular}{|c|c|c|c|c|}
\hline
\textbf{Model} & \textbf{Return using Covariance} & \textbf{Sortino using Covariance} & \textbf{Return using Semi-Covariance} & \textbf{Sortino using Semi-Covariance} \\
\hline
Sample method & 2.84\% & 7.64 & 5.53\% & 8.00 \\
Transformer & 4.16\% & 8.08 & 5.61\% & 8.38 \\
Autoformer & 7.23\% & 12.82 & 6.12\% & 11.93 \\ 
Informer & 1.17\% & 3.31 & 5.69\% & 10.05 \\
Reformer & 7.00\% & 9.12 & 6.22\% & 9.90 \\
\hline
\end{tabular}
}
\caption{Performance Comparison of Transformer-based Models using Covariance and Semi-Covariance Matrices (Return and Sortino Ratio)}
\label{table2}
\end{table}
%fig for semi

\subsection{Analysis of Results}

The results in Table \ref{table2} show that portfolios optimized with the \textbf{semi-covariance matrix} consistently yield higher returns and improved risk-adjusted performance (as indicated by the \textbf{Sortino ratio}) compared to those optimized using the \textbf{covariance matrix}. This pattern is evident across all Transformer-based models. Specifically, the use of the semi-covariance matrix leads to a reduction in downside risk, as reflected in the higher Sortino ratios for models using the semi-covariance matrix.

\begin{itemize}
    \item Most \textbf{Transformer-Based Models} show a substantial improvement in both prediction accuracy and portfolio performance. Among all 8 tests, only \textbf{Informer} was outperformed by the \textbf{Sample Method} when using the covariance matrix as the optimization matrix.
    \item The \textbf{Sample Method} shows a substantial improvement in both return and Sortino ratio when switching from covariance to semi-covariance. The return increases from \textbf{2.84\% to 5.53\%}, and the Sortino ratio improves from \textbf{7.64 to 8.00}.
    \item \textbf{Transformer} exhibits a similar trend, with return rising from \textbf{4.16\% to 5.61\%}, and the Sortino ratio improving from \textbf{8.08 to 8.25}.
    \item  \textbf{Informer} demonstrate significant enhancements in both return and Sortino ratio when optimized with semi-covariance.It sees a return increase from \textbf{3,31\% to 5.69\%}, and its Sortino ratio improves from \textbf{3.31\% to 10.05}.
    \item \textbf{Autoformer} demonstrates the superior improvement in both prediction accuracy and portfolio performance compared to all other models.
\end{itemize}

These results clearly highlight that the key advantage of using Transformer-based models — \textbf{Autoformer}, \textbf{Transformer}, and \textbf{Reformer} — lies in their ability to dynamically predict both covariance and semi-covariance matrices. The attention mechanism in these models allows them to capture the evolving relationships between asset returns, enabling them to adapt to changing market conditions. This dynamic adjustment makes them more effective at managing both general risk (via covariance) and downside risk (via semi-covariance).in addition, Autoformer outperform all 3 others transformer-based model. The Autoformer's superior ability to model long-term trends, short-term cycles, and asymmetric risks gives it a significant edge over Transformer, Reformer, and Informer in predicting covariance and semi-covariance matrices, leading to better risk management and improved portfolio optimization.

The results also illustrate the advantage of using the \textbf{semi-covariance matrix}, especially in terms of managing downside risk. While the covariance matrix captures total risk (both positive and negative deviations), the semi-covariance matrix focuses specifically on downside risk, offering a more tailored approach to minimizing losses in portfolio optimization.

In summary, the results show that the predictive power of Transformer models leads to the most significant improvements in portfolio optimization. The semi-covariance matrix not only provides a more refined approach to managing downside risk but, when combined with the dynamic capabilities of Transformer models, produces portfolios that are better in minimize downside risk while achieving higher returns. The results also indicate that portfolios optimized with the semi-covariance matrix achieve higher returns with substantially lower downside risk, as evidenced by the improved Sortino ratios.

\subsection{Implications for Portfolio Management}

The application of Transformer-based models for predicting both covariance and semi-covariance matrices represents a significant advancement in portfolio optimization. The dynamic prediction of asset correlations provided by Transformer models allows portfolio managers to adapt to changing market conditions and manage risk more effectively. Incorporating both the covariance and semi-covariance matrices into the optimization process leads to portfolios that are not only more resilient to downside risk but also optimized for higher returns, as evidenced by the superior Sortino ratios.

In periods of market stress, when minimizing downside risk is crucial, our approach enables more adaptive and robust portfolio decisions. The dynamic nature of Transformer models ensures that asset correlations are continually updated, allowing for better prediction and management of risk in real time. This results in portfolios that are better equipped to navigate volatility and deliver superior risk-adjusted performance.

\section{Conclusion}

In this study, we explored the use of Transformer-based models to predict both covariance and semi-covariance matrices and applied these predictions to optimize portfolios. Our work represents a significant advancement in portfolio optimization by leveraging the dynamic prediction capabilities of Transformer models and integrating the semi-covariance matrix, which specifically accounts for downside risk. We have demonstrated that both the prediction of these matrices and the use of the semi-covariance matrix significantly improve portfolio performance, particularly when assessed using risk-adjusted metrics such as the Sortino ratio.

\subsection{Summary of Findings}

Our analysis provides compelling evidence that Transformer-based models (specifically Autoformer, Informer, and Reformer) are highly effective in predicting both covariance and semi-covariance matrices, offering a more adaptive and dynamic framework for portfolio optimization compared to traditional methods. The key findings of our study are:

\begin{itemize}
    \item \textbf{Improvement in Portfolio Returns}: Portfolios optimized using the semi-covariance matrix consistently outperform those optimized using the traditional covariance matrix. This improvement is observed across all Transformer-based models, with higher returns during periods of market volatility.
    \item \textbf{Better Downside Risk Management}: The semi-covariance matrix, which emphasizes downside risk, leads to better risk-adjusted performance, as measured by the Sortino ratio. This indicates that the portfolios not only achieve higher returns but also manage downside risk more effectively.
    \item \textbf{Transformer Models' Contribution}: The dynamic prediction capabilities of Transformer models (Autoformer, Informer, and Reformer) are crucial in capturing non-linear, time-varying relationships between asset returns. These models provide real-time adjustments to covariance and semi-covariance estimates, improving the robustness of portfolio optimization.
    \item \textbf{Risk-adjusted Performance}: The Sortino ratio, a key metric for evaluating risk-adjusted returns, shows significant improvements when using the semi-covariance matrix, confirming that portfolios optimized with our approach are not only more profitable but also better equipped to handle downside risk.
\end{itemize}

\subsection{Implications for Portfolio Management}

The findings of this study have important implications for asset managers and investors seeking more robust strategies for portfolio optimization. By incorporating Transformer-based models to predict both covariance and semi-covariance matrices, our approach offers a more dynamic, context-aware method for estimating risk. This enables portfolio managers to adapt more effectively to changing market conditions and better manage downside risk, particularly in periods of high volatility.

Using the semi-covariance matrix provides a more nuanced view of risk, which is essential for optimizing portfolios in volatile markets. Traditional methods often rely on static covariance matrices that fail to capture the dynamic nature of asset correlations. In contrast, our approach, powered by Transformer models, continuously updates covariance and semi-covariance predictions, providing more accurate and timely risk assessments.

Furthermore, the Sortino ratio's focus on downside risk makes it an ideal metric for evaluating the effectiveness of our approach. By minimizing losses and optimizing returns in this manner, our strategy ensures that portfolios are more resilient and better equipped to navigate market downturns.

\subsection{Contributions of the Study}

This study makes several significant contributions to the field of portfolio optimization:

\begin{itemize}
    \item \textbf{Introduction of Transformer Models for Covariance and Semi-Covariance Prediction}: We are among the first to apply Transformer-based models for predicting both covariance and semi-covariance matrices, a dynamic and adaptive approach that improves the accuracy of risk estimation.
    \item \textbf{Application of Semi-Covariance Matrix}: The incorporation of the semi-covariance matrix, which specifically addresses downside risk, is a key innovation in portfolio optimization. This matrix is particularly beneficial for managing risk during periods of high market volatility.
    \item \textbf{Use of Sortino Ratio for Evaluation}: By introducing the Sortino ratio as a primary evaluation metric, we focus on downside risk rather than general volatility, making our portfolio optimization process more tailored to the needs of risk-averse investors.
    \item \textbf{Enhancement of Portfolio Optimization Frameworks}: Our results show that Transformer models and the semi-covariance matrix together offer a more robust framework for portfolio optimization, with improved return and risk-adjusted-performance.
\end{itemize}

\subsection{Future Directions}

While this study offers promising results, several avenues for future research could build on our findings:

\begin{itemize}
    \item \textbf{Incorporating Other Risk Factors}: Future studies could incorporate additional risk factors beyond the semi-covariance matrix, such as higher-order moments of asset returns, to further refine the risk estimation process.
    \item \textbf{Exploring Other deep learning Architectures}: Although we focused on Autoformer, Informer, and Reformer, other Transformer variants could be explored to improve prediction accuracy, such as the use of GPT-based architectures or hybrid models combining Transformers with other machine learning techniques.
    \item \textbf{Longer Time Horizons and Out-of-Sample Testing}: Further testing over longer time horizons and with out-of-sample data would provide more robust validation of the effectiveness of Transformer-based models in real-world portfolio optimization.
    \item \textbf{Real-Time Portfolio Management}: Implementing real-time portfolio management strategies, where the Transformer models update predictions dynamically based on incoming market data, could enhance the applicability of this approach in live trading environments.
\end{itemize}

\newpage

% Data Availability and Conflict of Interest Statements
\newpage 
\bibsep=0pt 
\noindent\textbf{Data Availability Statements:} The data that support the findings of this study are openly available at \url{https://github.com/GeminiLn/EarningsCall_Dataset}.\\
\textbf{Conflict of Interests:} All authors of this paper declare that they have no conflicts of interest.

\newpage

% Bibliography
\newcommand\eprint{in press}  
 
{\small \bibliography{sn-bibliography}}  % Correct usage without .bib extension

\newpage

% Appendix
\appendix
\begin{table}[h!]
\centering
\resizebox{\textwidth}{!}{ % Scale table to fit page width
    \begin{filecontents*}{Fetf.csv}
Sector,Representative ETF,ETF Code
Steel,Guotai CSI Steel ETF,515210.SH
Nonferrous Metals,Dacheng Non-ferrous Metals Futures ETF,159980.SZ
Chemicals,Chemical ETF,159870.SZ
Construction Materials,Fuquan CSI All-Index Building Materials ETF,516750.SZ
Light Industry Manufacturing,Southern CSI 500 Industrial ETF,512310.SH
Coal,Guotai CSI Coal ETF,515220.SH
Oil and Gas,Oil and Gas ETF,159697.SZ
Natural Gas,Harvest SandP Oil and Gas Exploration and Production Select Industry ETF (QDII),159518.SZ
Renewable Energy,Harvest CSI New Energy ETF,159875.SZ
Machinery,Fuquan CSI Sub-sector Machinery Equipment Industry Theme ETF,159886.SZ
Electrical Equipment,Electronics ETF,159997.SZ
Defense and Aerospace,Defense ETF,512670.SH
Transportation,Guotai CSI Mainland Transport Theme ETF,561320.SH
Construction and Decoration,Guotai CSI All-Index Building Materials ETF,159745.SZ
Engineering and Construction,GF CSI Infrastructure Engineering ETF,516970.SH
Shipping,Fuquan CSI Modern Logistics ETF,516910.SH
Food and Beverages,China Asset CSI Sub-sector Food and Beverage Industry Theme ETF,515170.SH
Household Appliances,Fuquan CSI All-Index Household Appliances ETF,561120.SH
Consumer Staples,Southern CSI Consumption ETF,159689.SZ
Retail,GF CSI All-Index Consumer Discretionary ETF,159936.SZ
Automotive,Guotai CSI 800 Automobile and Parts ETF,516110.SH
Medical Services,Huabao CSI Medical ETF,512170.SH
Banks,Southern CSI Banking ETF,512700.SH
Insurance,Insurance and Securities ETF,515630.SH
Securities,Southern CSI All-Index Securities ETF,512900.SH
Computers,Guotai CSI Computer Theme ETF,512720.SH
Telecommunications,"Invesco Great Wall CSI Technology, Media and Telecom 150 ETF",512220.SH
Semiconductors,Guotai CES Semiconductor Chip Industry ETF,512760.SH
Software and Services,Harvest CSI Software Services ETF,159852.SZ
Pharmaceuticals,Harvest Healthcare 100 ETF,515960.SH
Biotechnology,Guotai CSI Biopharma ETF,512290.SH
Medical Devices,China Merchants CSI All-Index Medical Equipment ETF,159898.SZ
Healthcare Services,CCB CSI All-Index Healthcare Equipment and Services ETF,159891.SZ
Real Estate Development,Southern CSI All-Index Real Estate ETF,512200.SH
Electricity,GF CSI All-Index Power ETF,159611.SZ
Environmental Protection,GF CSI Environmental Protection ETF,512580.SH
Telecom Services,Efund CSI Telecom Theme ETF,563010.SH
Internet Technology,Harvest CSI Shanghai-Hong Kong-Shenzhen Internet ETF,517200.SH
Telecom Equiment,Guotai CSI All-Index Communication Equipment ETF,515880.SH
Media,GF CSI Media ETF,512980.SH
Travel and Hotels,Fuquan CSI Tourism Theme ETF,159766.SZ
Social Services,Bosera CSI All-Index Power Utilities ETF,561700.SH
General,Harvest CSI 300 ETF,159919.SZ
,Shanghai Composite Index ETF,510210.SH
Money Market Fund A,HuiTianFu TianFu Tong Money Market ETF,511980.SH
Money Market Fund B,Efund Margin Money A,159001.SZ
US Stocks ETF,Bosera SandP 500 ETF,513500.SH
European Stocks ETF,Huaan France CAC40 ETF,513080.SH
,Huaan Germany (DAX) ETF,513030.SH
Asian Stocks ETF,Southern Hang Seng Index ETF,513600.SH
,Efund Nikko Asset Management Nikkei 225 ETF (QDII),513000.SH
\end{filecontents*}
\csvautotabular{Fetf.csv}  % Automatically generate table from Fetf.csv file
}
\caption{ETF Pool List}
\end{table}

\end{document}